\documentclass[showpacs,aps,prb,twocolumn,groupedaddress,amsmath,amssymb]{revtex4}
\usepackage{graphicx}

\newcommand{\tr}{\ensuremath{\operatorname{tr}}}
\newcommand{\Tord}[1][\tau]{\ensuremath{\operatorname{T}_{#1}}}
\newcommand{\Tfor}{\Tord[\leftarrow]}
\newcommand{\Tback}{\Tord[\rightarrow]}
\newcommand{\ev}[1]{\ensuremath{\langle #1 \rangle}}
\renewcommand{\Re}{\ensuremath{\operatorname{Re}}}
\renewcommand{\Im}{\ensuremath{\operatorname{Im}}}
\newcommand{\sgn}{\ensuremath{\operatorname{sgn}}}
\newcommand{\Eone}{\ensuremath{\operatorname{E_1}}}
\newcommand{\GammaS}{\ensuremath{\Gamma}}
\newcommand{\rdef}{\ensuremath{\mathrel{\mathop:}=}}
\newcommand{\abs}[1]{\ensuremath{|#1|}}
\newcommand{\braketop}[3]{\ensuremath{\langle #1|#2|#3\rangle}}
\newcommand{\id}{\ensuremath{\openone}}

\begin{document}
	\title{Semi--classical dynamics of nano--electromechanical systems}
	
	\author{R. Hussein}
	\email[]{robert@itp.physik.tu-berlin.de}
	\author{A. Metelmann}
	\author{P. Zedler}
	\author{T. Brandes}
	\affiliation{Institut f\"ur Theoretische Physik, TU Berlin, Hardenbergstr. 36, Berlin D-10623, Germany}
	\date{\today}
	
	\begin{abstract}
		We investigate the dynamics of a single phonon (oscillator) mode linearly coupled to an electronic few--level system in contact with external particle reservoirs (leads). A stationary electronic current through the system generates non--trivial dynamical behaviour of the oscillator. Using Feynman--Vernon influence functional theory, we derive a Langevin equation for the oscillator trajectory that is non--perturbative in the system--leads coupling and from which we extract effective oscillator potentials and friction coefficients. For the two simplest cases of a single and two coupled electronic levels, we discuss various regimes of the oscillator dynamics. 
	\end{abstract}
	
	\maketitle

	\section{Introduction}
	Nano--electromechanical systems (NEMS) are test--beds for the observation of fundamental quantum behaviour of objects which are huge on the scale of individual atoms. For example, recent experiments \cite{OConnelletal2010}
	have allowed a detailed study and control of single phonons by cooling a macroscopic resonator mode close to its ground state and coupling it to single electronic degrees of freedom.
	
	One fascinating aspect of NEMS is their conceptual simplicity that nevertheless can give rise to highly complex
	physics, and the links that can be established to other fields such as molecular electronics or optomechanics \cite{Groetal09,MG09}. One of the  challenges are the details of the oscillator--electron coupling, i.e. in the language of measurement theory to understand, utilize \cite{Naiketal06} and control \cite{CMJ08,AB08} the `back--action' effects of the detector (e.g., superconducting single--electron transistors \cite{LaHayeetal09,Blen04}) onto the oscillator.
	
	In many cases, even if no further approximations in simple theoretical models are made,
	the coupling to external electronic reservoirs that link the detector to the 
	outer world is treated perturbatively, i.e., in the framework of 
	(quantum) Master equations. This has turned out to be a highly successful 
	approach, in particular to describe such various  systems as 
	NEMS coupled to single electron transistors (SETS) \cite{Rodrigues2007,Rodrigues2005,Armour2004},
	Franck--Condon blockades \cite{KochAll}
	in transport through molecules with strong electron--phonon coupling, 
	or quantum shuttles\cite{Fedorets04,Armour02,Novotny03,Novotny04}. From the theory
	of electronic transport through nanostructures \cite{Brandes2005}, 
	however, it is known that such approximations
	usually are reliable only in the limit of high external voltage bias,
	where non--Markovian effects \cite{BraggioFlindt}        
	due to quantum coherences between the external 
	reservoirs and the electronic system (SET, quantum dot etc.) can be neglected. 
	It is therefore desirable to develop tools that allow a description of NEMS 
	beyond the Master equation regime (weak electron--leads coupling) and at the same 
	time are not merely perturbative 
	in the coupling of the oscillator to the electronic environment \cite{Mitra04}.
	
	In the past, the coupling of electrons to a single bosonic mode
	has been solved exactly for the case where only
	one single electron is present \cite{Glazman88,Wingreen89}, i.e. in an empty band 
	approximation. The inclusion of  Fermi sea reservoirs at different chemical potential
	transforms this into a difficult many--body problem out of equilibrium, and 
	approximations are necessary  \cite{Mitra04,PhysRevB.68.205323,PhysRevB.70.245306,Mozyrsky2006}.
	
	Our approach in this paper is to combine exact solutions of the electronic system
	with a semi--classical expansion, together with an adiabatic approximation 
	for the oscillator dynamics within the Feynman--Vernon
	influence functional (double path integral) theory \cite{Feynman1963}. 
	We revise this method, which has first been used for simple 
	NEMS models by Mozyrsky and co--workers \cite{Mozyrsky2004}, and extend it to allow 
	for the description of 
	a relatively large class of non--equilibrium electronic environments. 
	The key idea is a systematic  expansion 
	around the classical path 
	in order to obtain a Langevin equation for the oscillator. 
	Already at the simplest level of this approximation (neglecting quadratic fluctuations
	around the diagonal path in the reduced density matrix of the oscillator), the 
	coupling to the electronic non--equilibrium environment gives rise
	to non--trivial effects such as effective oscillator potentials 
	and non--linear friction coefficients leading to both positive and
	negative damping \cite{Bennett2006}. 
	Gaussian fluctuations around the classical path
	are built into the theoretical description, 
	but they have to be evaluated by numerical solutions
	of the underlying Langevin equations which is not done 
	in this paper. 
	
	We compare two non--interacting electronic `quantum dot' models with one and two levels between source and drain reservoirs: a single dot, and two dots in series. The oscillator couples linearly to the dot occupation (single dot) or to the occupation difference (double dot). One particular feature of the double dot case (where quantum superpositions of the electrons become important) is the occurence of limit cycles in phase space caused by a negative damping.
	
	The paper is organized as follows: 
	after introducing the path integral formalism with a generic model in Sec. II,
	we present the single dot case in Sec. III and the double dot case in Sec. IV.
	Detailed derivations of the important formulae can be found in the 
	appendices.
	\section{Generic model}
	A large class of NEMS can be described as a composition of an electronic system $\mathcal H_{\rm e}$, a mechanical system $\mathcal H_{\rm osc}$ and a linear coupling between the two. Thus we set up a generic Hamiltonian $\mathcal H_{\rm  gen}$ by
	\begin{align}
	\mathcal H_{\rm gen} &= \mathcal H_{\rm e} + \mathcal H_{\rm osc} - \hat F \hat q, \label{eq.:H_gen}\\
	\mathcal H_{\rm osc} &=  \frac{1}{2 m} \hat p^2 + V_{\rm osc}(\hat q).
	\end{align}
	Here, $\mathcal H_{\rm osc}$ describes a single oscillator with $\hat p$ (momentum) and $\hat q$ (position) operators. The oscillator mode is confined in a potential $V_{\rm osc}(q)$, $\hat F$ denotes an electronic force operator, and $m$ labels the oscillator mass. Whithin this paper the reduced Planck constant is set to one ($\hbar=1$).
	
	Our generic model does not include an additional oscillator damping mechanism . 
	In the usual Master equation treatment  of NEMS, Lindblad-form damping due to external degrees of freedoms 
	is included phenomenologically. In the path integral 
	formalism used here, such degrees of freedom can be easily included at least for linear or weak coupling to the oscillator. 
	In order to elucidate the effect of the electronic environment that we treat in all orders in the coupling to external 
	electronic reservoirs (contained in $\mathcal H_{\rm e}$), we choose not to include additional damping terms in our model here.
	
	\subsection{Stochastic equation of motion}
	We describe the oscillator dynamics by the reduced density matrix of the oscillator in position representation $\rho_{\rm osc}(q,q',t)=\braketop{q}{\rho_{\rm osc}(t)}{q'}$, for which
	we derive a semiclassical equation of motion for the oscillator position by using Feynman--Vernon influence functional theory similar to Mozyrsky and co-workers\cite{Mozyrsky2006}.
	Assuming that the total density matrix $\chi(t)$ factorises at the initial time $t_0$ into a system and a bath part $\chi(t_0)=\rho_{\rm{osc}}(t_0)\otimes\rho_{\rm B}$, the
	propagation of the reduced oscillator density $\braketop{q}{\rho_{\rm osc}(t)}{q'}$ at time $t$ is given by a double path integral\cite{Schulman2005,Weiss2008}, cf.~appendix~\ref{sec.:gen_mod}.
	A transformation to center--of--mass and relative coordinates
	\begin{align}
	x_t &= \frac{q_t + q'_t}{2}, \qquad y_t = q_t - q'_t \label{eq.:gen_xy_expansion}
	\end{align}
	has the notion to detach the classical trajectory $x_t$ from the quantum mechanical deviations. Within a Born--Oppenheimer approximation, change of variables allows us to study a slow oscillator by an adiabatic approximation of the classical trajectory
	\begin{align}
	x_t &\approx x_0+t\dot x_0, \label{eq.:gen_adiabatic_approx}
	\end{align}
	cf.~appendix~\ref{sec.:gen_mod}. In this approach, the typical timescale of the oscillator movement is slow compared to the electronic transition rates. In the subsequent, when having introduced the angular oscillator frequency by $\omega_0$ and electron transition rates by $\Gamma_{\rm L}$, $\Gamma_{\rm R}$ we have to satisfy the condition $\omega_0\ll\Gamma_{\rm L}$, $\Gamma_{\rm R}$.
	
	In the next step, we derive a stochastic equation of motion for the classical trajectory, taking into account the propagation of the initial reduced density matrix, cf.~appendix~\ref{sec.:gen_mod}. 
	The key step here is a cluster expansion to quadratic order in the off-diagonal path $y_t$ that describes the Gaussian fluctuations around the classical 
	oscillator trajectory, where the fluctuations are determined by the properties of the non-equilibrum environment.
	To achieve a self--consistent equation of motion, we then re-insert the full time--dependence of the fixed classical trajectory in accordance with the adiabatic approximation and end up with the Langevin equation
		
	\begin{align}
	&m \ddot x_t  +V'_{\rm osc}(x_t) -\ev{\tilde F[x](t)}
		+\dot x_t A[x](t) = \xi_t\label{eq.:gen_model_langevin}.
	\end{align}
	Here the interaction picture of the electronic operator is given by 
	$\tilde F[x](t) = \exp[i(\mathcal H_{\rm e}-\hat F x)t]\hat F\exp[-i(\mathcal H_{\rm e}-\hat F x)t]$,
	and $\xi_t$ is a stochastic force with zero mean and the correlation function
	\begin{align}
	\ev{\xi_t\xi_{t'}} &=
		2\Re\ev{\delta\tilde F[x](t)\delta\tilde F[x](t')}.
	\end{align}
	The fluctuation of the electronic operator is defined by $\delta\tilde F[x](t)=\tilde F[x](t)-\ev{\tilde F[x](t)}$.
	The friction $A[x]$ is given by 
	\begin{align}
	A[x](t) &= 2\int_{t_0}^{t}dt'\;t'\;\Im\ev{\delta\tilde F[x](t)\delta\tilde F[x](t')}.
	\end{align}
	In the following we choose $t_0=-\infty$ as initial time. For the specific cases of single  and double dots, we checked that the upper integration boundary can be extended to infinity.

	\section{Anderson--Holstein model (AHM)}
	The AHM combines a single bosonic mode
	with a simple electronic transport system.
	We describe the bosonic part in first quantisation as
	\begin{equation}\label{eq.:H_osc}
		\mathcal H_{\rm osc}=\frac 1{2m} \hat p^2 +\frac12 m\omega_0^2 \hat x^2
	\end{equation}
	with the bosonic position and momentum operators $\hat x$ and $\hat p$,
	the phonon frequency $\omega_0$
	and the phonon mass $m$.
	The oscillator length is defined by $l_0\equiv[m\omega_0]^{1/2}$.
	The electronic part is a single dot level
	confined between two leads:
	\begin{equation}
		\mathcal H_{\rm e}=\varepsilon_{\rm d}\hat d^\dag \hat d 
		+\sum_{k\alpha}\varepsilon_{k\alpha}\hat c_{k\alpha}^\dag\hat c_{k\alpha}
		+\sum_{k\alpha}\big(V_{k\alpha} \hat c_{k\alpha}^\dag \hat d +\text{H. c.}\big)\nonumber.\\
	\end{equation}
	The dot level has energy $\varepsilon_{\rm d}$
	and creation/annihilation operators $d^\dag$/$d$.
	The operators $\hat c^\dag_{k\alpha}$/$\hat c_{k\alpha}$
	create/anihilate electrons with momentum $k$
	and energy $\varepsilon_{ka}$
	in a free electron gas in the left ($\alpha=\rm L$)
	or right ($\alpha=\rm R$) lead.
	Electronic transitions are possible with amplitude $V_{ka}$
	between the dot and a state in the lead.
	Between the two subsystems there is a simple linear coupling
	with coupling constant $\lambda$,
	such that the total Hamiltonian reads
	\begin{align}
	\mathcal H_{\rm AHM}&=\mathcal H_{\rm e}+ \mathcal H_{\rm osc}-\lambda \hat d^\dag \hat d\hat x.
	\end{align}
	For convenience we introduce the dimensionless coupling constant:
	\begin{align}
	g=\frac{\lambda}{m\omega_0^2l_0}\label{eq.:SET_coupling}.
	\end{align}\\
	Here, we regard spinless electrons. Note that the generalization to a model including both spin directions requires an onsite interaction term like $U\hat n_\uparrow\hat n_\downarrow$. In the following, we only consider non--interacting electrons.
	A physical realization of the model discussed here would correspond to either spin polarized electrons or a coupling to the oscillator that effects only electrons of one certain spin direction.
	
	\begin{figure}[t]
		\centering
		\includegraphics[width=\linewidth]{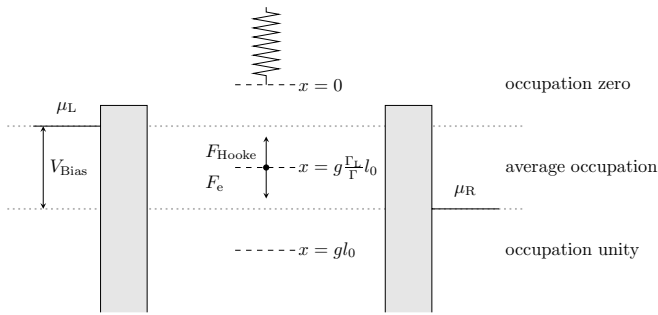}		\caption{\label{fig.:SET_force_balance}
		Sketch of the AHM model. Two leads (grey bars) with chemical potentials $\mu_{\rm L}$, $\mu_{\rm R}=\mu_{\rm L}-eV_{\rm Bias}$ embed the dot level which couples to an oscillator. Here $V_{\rm Bias}$ is the bias voltage and $e$ the electron charge. The level energy $\varepsilon_{\rm d}$ is shifted by the oscillator position $x$, the shifted level energy $\varepsilon_x = \varepsilon_{\rm d} -\lambda x$ is stationary, if the oscillator force $F_{\rm Hooke}=m\omega_0^2 x$ and the electron force $F_{\rm e} =\lambda\ev{\hat n[x]}$ are in balance. %
		Here $m$ and $\omega_0$ describe the oscillator mass and angular frequency, and $\lambda$ (g) denotes the coupling constant, cf.~eq.\eqref{eq.:SET_coupling}. For sufficiently small tunneling rates $\Gamma_{\rm L}$, $\Gamma_{\rm R}$ the stable stationary level energies (dashed lines) are located at $x=0,gl_0\Gamma_{\rm L}/\Gamma,gl_0$. The instable points of $\varepsilon_{x}$ (dotted lines) are located around the chemical potentials.%
		}
			
	\end{figure}
	
	\subsection{Langevin equation}
	When applying our generic form~\eqref{eq.:H_gen} to the AHM, we obtain the Langevin equation
	\begin{align}
	m\ddot x_t -F_ {\rm eff}(x_t )+\dot x_t A[x](t) &= \xi_t \label{eq.:SET_Langevin_eq}
	\end{align}
	with the effective force $F_{\rm eff}$ and the friction $A[x]$:
	\begin{align}
	F_ {\rm eff}(x_t ) &= 
		-m\omega_0^2 x_t + \lambda \ev{\tilde n(t)}
	= -\frac{\omega_0}{l_0}\bigg[
			\frac{x}{l_0}- g \ev{\tilde n(t)}
		\bigg],\nonumber\\
	A[x](t) &= 2 \lambda^2 \int_{t_0}^{t}dt'\;t'
		\Im\ev{\delta\tilde n(t)\delta\tilde n(t')}.
	\end{align}
	The effective force has two contributions; a term proportional to the elongation (Hooke's law) and the electron force which is proportional to the dot occupation; which only contributes if the dot is occupied. The friction term $x_t A[x]$ results from stochastic electron jumps between the leads and the dot.
	
	\begin{figure}[t]
		\centering
		\includegraphics[width=\linewidth,clip]{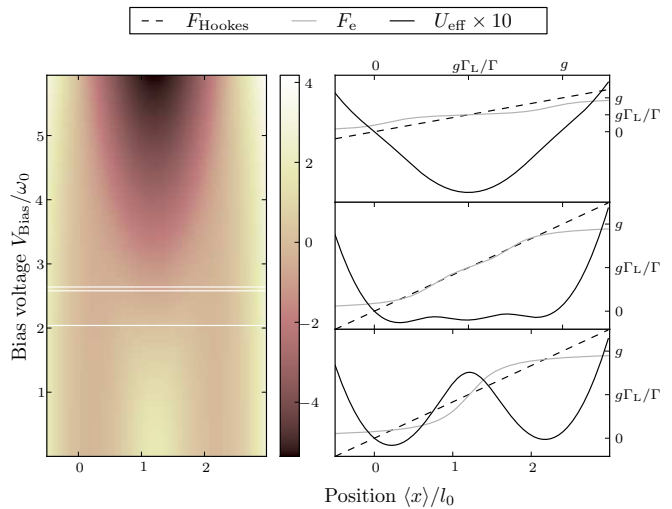}		\caption{\label{fig.:SET_density_composed_sliced}
		\textsc{left}) Density plot of the effective oscillator potential $U_{\rm eff}$ as a function of oscillator position $\ev{x}$ and $V_{\rm Bias}$ in units of $\omega_0$ and $l_0$ with the parameters $\Gamma_{\rm L,R}/\omega_0=0.7$, $\varepsilon_{\rm d}/\omega_0=2.9$, $\beta\omega_0=10$ and $g=2.4$.
		For increasing bias voltage, the effective potential shows two, three, two and one minima. This different regions are separated by white lines at $V_{\rm Bias}/\omega_0= 2.04/2.58/2.64$. 
		\textsc{Right}) $U_{\rm eff}$ and the two contributions $F_{\rm Hooke}$, $F_{\rm e}$ to the force $F_{\rm eff}$ at bias voltages (symmetric choice) $V_{\rm Bias}/\omega_0=0.0/2.3/5.0$. $U_{\rm eff}$ exhibits extrema where the two force contributions (grey and dashed line) are in balance. The width of the center plateau in the electronic force contribution $F_{\rm e} \varpropto \ev{n}$ grows with increasing bias voltage $V_{\rm Bias}$, cf.~Fig.~\eqref{fig.:SET_force_balance} for the intermediate occupation $\ev{n}$. For sufficiently high bias voltage only one minimum remains. %
		}
		
	\end{figure}
	\noindent For finite bias voltage, figure~\ref{fig.:SET_force_balance} shows the positions of the dot energy level and the points of instable balance, which result from the balance of both forces.

	The effective force and therewith the oscillator potential are determined by the dot occupation $n(t)$, whereas the friction and the stochastic force correlation 
	$\ev{\xi_{t}\xi_{t'}} = \lambda^22\Re\ev{\delta\tilde n(t)\delta\tilde n(t')}$ depend on the imaginary/real part of the dot correlation function.
	The dot correlation function in terms of the lesser and greater Green's function (which are derived in appendix~\ref{sec.:AHM_Corr})  reads 
	\begin{equation}
	\ev{\delta\tilde n(t)\delta\tilde n(0)}= G^<(-t)G^>(t).
	\end{equation}

	\subsection{Effective potential} \label{subsec.:AHM_Eff_pot}
	\begin{figure*}[t]
		\centering
		\includegraphics[width=0.9\linewidth]{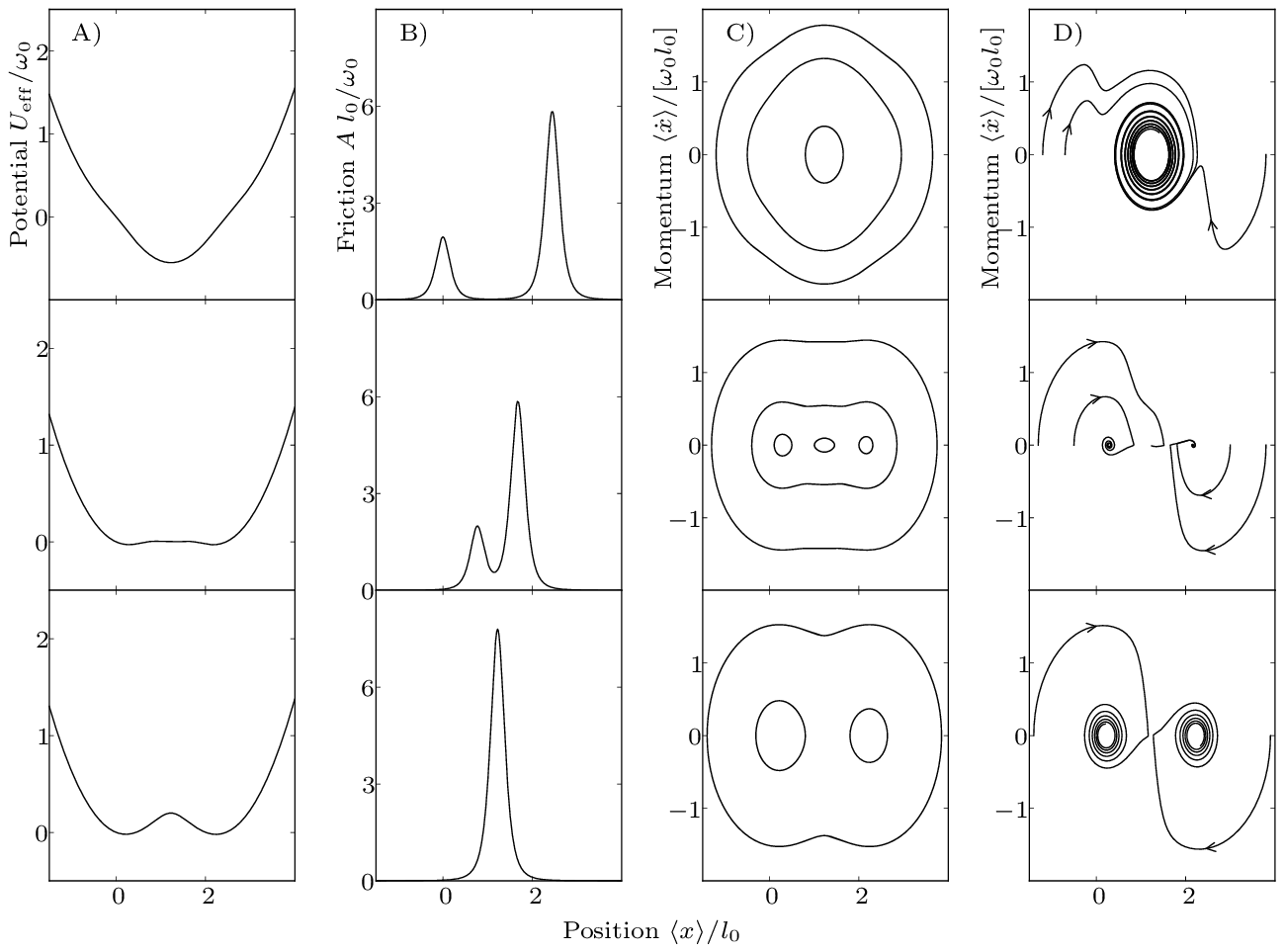}		\caption{\label{fig.:SET_phase_space_comparison}
		Effective oscillator potentials (A), frictions $A(x)$ (B) and the phase space portraits without (C) and with (D) friction in units of $\omega_0$ and $l_0$ with the parameters $\Gamma_{\rm L,R}/\omega_0=0.7$, $\varepsilon_{\rm d}/\omega_0=3.0$ and $g=2.45$ at zero temperature. From the bottom to the top the values of the bias voltage (symmetric choice) read $V_{\rm Bias}/\omega_0=0.0/2.2/6.0$ in correspondence to the three regions in Fig.~\ref{fig.:SET_density_composed_sliced} (\textsc{Right}). The peaks of the friction are located at $x=[\varepsilon_{\rm d} \mp V_{\rm Bias}/2]/\lambda$ where the shifted level energies $\varepsilon_x$ are in resonance with the chemical potentials $\mu_{\rm L,R}$, cf.~Fig.~\ref{fig.:SET_force_balance}.
		}
			
	\end{figure*}
	The occupation of the dot is calculated in an adiabatic approach with the help of the lesser Green's function 
	$G^<(\omega)$, cf. appendix \ref{sec.:AHM_Corr}, where we assume constant tunneling rates $\Gamma_\alpha$. 
	For finite temperatures, the dot occupation reads
	\begin{align}
	\ev{\tilde n(t)} &= 
		-i\frac{1}{2\pi} \int d\omega\; G^<(\omega) \nonumber\\
	&= \frac12 -\frac{1}{\pi}\sum_\alpha\frac{\Gamma_\alpha}{\Gamma}
		\Im\Psi\bigg(\frac12 +\frac{\beta\Gamma}{4\pi} 
		+i\frac{\beta(\varepsilon_x-\mu_\alpha)}{2\pi}\bigg),
	\end{align}
	whereby $\beta$ stands for the inverse temperature, $\varepsilon_x$ is a short notation 
	for $\varepsilon_{\rm d}-\lambda\hat x$, $\mu_\alpha$ denotes the chemical potetials and $\Gamma\equiv\Gamma_{\rm L} + \Gamma_{\rm R}$. $\Psi$ designates the Digamma function. By integration we obtain the effective potential
	\begin{align}
	&U_{\rm eff}(x) = -\int_0^x dx'\; F_{\rm eff}(x') =\\
	&\frac{2}{\beta}\sum_\alpha\frac{\Gamma_\alpha}{\Gamma}\Re\bigg[
		\ln\GammaS\bigg(
			\xi+i\frac{\beta(\varepsilon_{\rm d}-\lambda x-\mu_\alpha)}{2\pi}
		\bigg)\nonumber\\
	&-\ln\GammaS\bigg( 
			\xi+i\frac{\beta(\varepsilon_{\rm d}-\mu_\alpha)}{2\pi}
		\bigg)
		\bigg]\bigg|_{\xi=\frac12 +\frac{\beta\Gamma}{4\pi}}\;
		+\frac12 \frac{x}{l_0}\bigg[\frac{x}{l_0}-g\bigg]\omega_0,
	\end{align}
	with $\GammaS(\cdot)$ denoting the Gamma function.
	In the absence of friction the dynamics of the oscillator is determined by the effective potential $U_{\rm eff}(x)$. 
	The high temperature case is of minor interest, because the temperature washes out the structures of $U_{\rm eff}(x)$.
	
	In Fig.~\ref{fig.:SET_density_composed_sliced} we present features of the effective potential $U_{\rm eff}(x)$ at zero temperature. 
	$U_{\rm eff}(x)$ has minima when the effective force $F_{\rm eff}(x)$ is zero. This is the case when the oscillator force and the electron force are in balance. We plot the two contributions to the effective force: the force $F_{\rm Hooke}$ is proportional to the displacement $\ev{x}$, the contribution $F_{\rm e}$ results from the dot occupation (scaled with $g$) and has two steps at 
	\begin{align}
	\frac{x_\alpha}{l_0} &= \frac{1}{g}\bigg[\frac{\varepsilon_{\rm d}}{\omega_0} -\frac{\mu_\alpha}{\omega_0}\bigg],\qquad\alpha\in\{\rm  L,\; R\}.
	\end{align}
	In the upper part of Fig.~\ref{fig.:SET_density_composed_sliced} we easily recognize that for large bias voltage there will be only one minimum at $x/l_0=g\Gamma_{\rm L}/\Gamma$. By changing the bias voltage or the coupling strength, we can reach situations where two minima at $x/l_0=0$ and at $x/l_0=g$ are added, like in the middle part of Fig.~\ref{fig.:SET_density_composed_sliced} and also situations, where only the two minima at the side remain and the one in the middle vanishes, like in the bottom part of the~\ref{fig.:SET_density_composed_sliced} (this agrees with Mozyrsky et~al.\cite{Mozyrsky2006}). Increasing bias voltage shifts the steps in $F_{\rm e}$ apart, whereas increasing the coupling constant minimises the distance. The positions of the minima/steps are exact in the zero rate limit ($\Gamma_{\rm L}\rightarrow0$, $\Gamma_{\rm R}\rightarrow0$) where the averaged occupation is step--like. For finite rates, the steps smoothes out.
	
	The form of the oscillator potential already hints towards the phase space spanned by position $\ev{x}$ and 
	momentum  $m\ev{\dot x}$. In the absence of friction, the minima of the potential correspond to the fixed points of the oscillator motion.

	\subsection{Phase space portrait}
	The classical trajectory $\ev {x_t}$ is obtained by neglecting fluctuations due to the stochastic force $\xi_t$ in \eqref{eq.:SET_Langevin_eq}; therefore the equation of motion reads
	\begin{align}
	m\ev{\ddot x_t} + \ev{\dot x_t} A[\ev{x}](t)-F_{\rm eff}(\ev{x_t}) \label{eq.:stochastical_eq_of_motion_cl} = 0.
	\end{align}
	In Fig.~\ref{fig.:SET_phase_space_comparison} we show the effective potential
	and the friction, as well as the solutions of the classical
	equation of motion in phase space without and with friction.
	The rows correspond to three different voltages leading to
	the three typical cases with one,
	three and two minima of the potential
	(the latter case is investigated in \cite{Mozyrsky2006}).
	The initial conditions are chosen for each phase diagram separately
	in order to make the characteristic shapes visible.
	The phase diagrams without friction follow directly from the shape
	of the effective potential.
	The trajectories including the friction follow the ones without friction
	for a while until a position with a peak in the friction is reached
	that produces a kink-like damping feature.
	The friction causing the kinks is maximal, when the shifted energy level $\varepsilon_x$ is in resonance with the chemical potentials $\mu_{\rm L,R}=\pm V_{\rm Bias}/2$ (Fig.~\ref{fig.:SET_force_balance}), because at these points the average occupation switches and the electronic fluctuations and therewith the friction itself is very large.
	
	Between its peaks the damping does not completely vanish
	at finite bias, so all phase space trajectories
	end up in spirals and reach stable fixed points after infinite time
	(we have terminated all phase trajectories after the same time).
	The position of the peaks (instable fixed points) separates the stable fixed points.
	In the two minima case ($V_{\rm Bias}/\omega_0=0.0$ in Fig.~\ref{fig.:SET_phase_space_comparison}) the left fixed point corresponds to the zero occupied electronic level in Fig.~\ref{fig.:SET_force_balance} and the right fixed point to the fully occupied level. For increasing bias voltage, the friction splits and we obtain a third fixed point in between, corresponding to the average occupied state. For sufficiently high bias voltage (large transport window) only the average occupied state survives.
	
	Apart from the stable fixed points we also observe saddle points
	that repell the trajectories near the kinks where the damping is large.
	They result from the kinks in the dot occupation
	(Fig.~\ref{fig.:SET_density_composed_sliced}, right) and correspond
	to the dotted lines in Fig.~\ref{fig.:SET_force_balance}
	and the bright V-structure in
	Fig.~\ref{fig.:SET_density_composed_sliced}, left.
	At large bias the saddle points move out of the range of allowed
	positions, the same happens to the peaks in the friction.
	At infinite bias the effective potential becomes a simple parabola
	and the friction vanishes completely
	(this can be checked analytically).

	\begin{figure*}[t]
		\centering
		\includegraphics[width=0.9\linewidth]{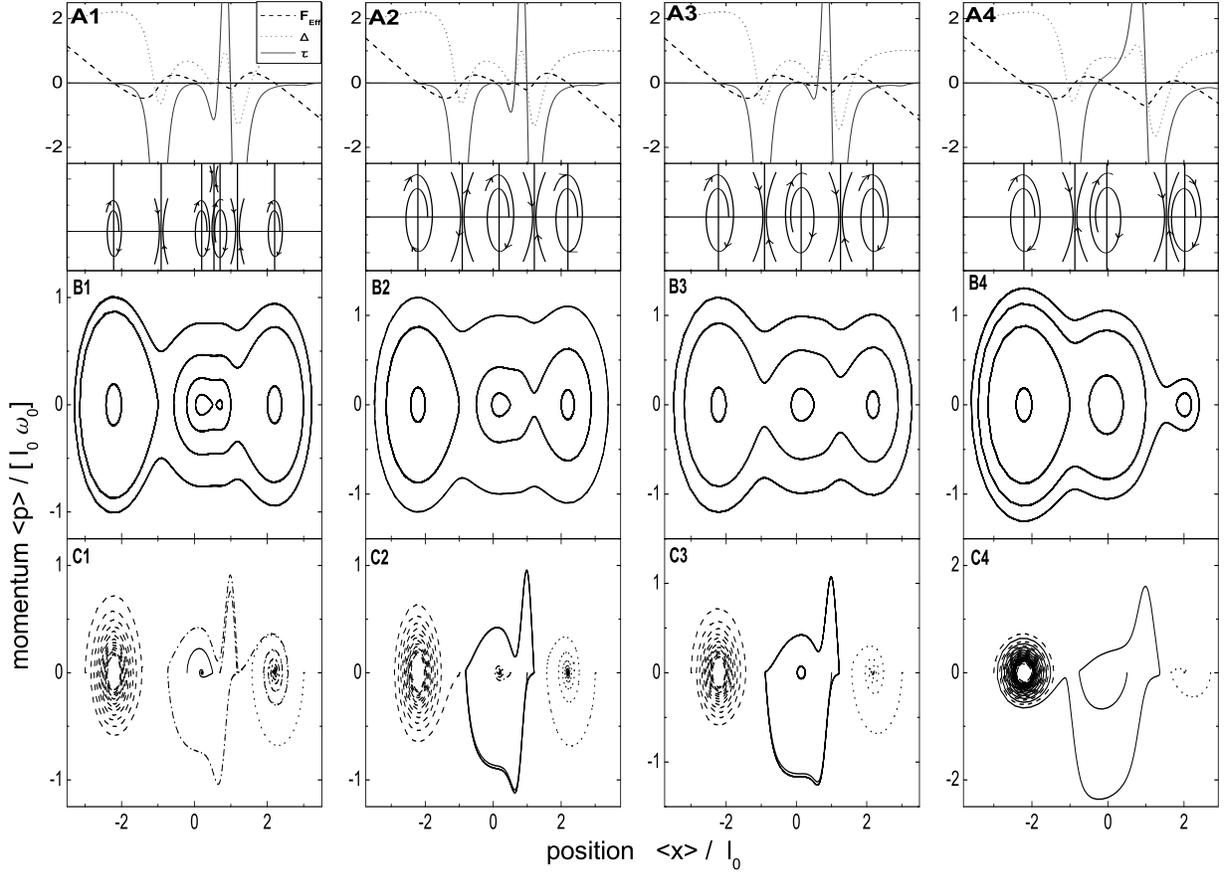}		\caption{\label{fig.:DQD_phase_space_comparison}
		Examination of the DQD system for various tunnel couplings $T_{c}$, increasing from left to right. 
		Row A shows the results of a fixed point analysis. $\Delta$ corresponds to the determinant and $\tau$ to 
		the trace of the Jacobian matrix. Rows B and C display phace space portraits without and with friction.
		Explicit parameters are $\abs{T_{c}}^2 = 0.4; 0.43; 0.49; 1.0  \ \omega_{0}^2$.
		We used equal rates $\Gamma_{\rm L}=\Gamma_{\rm R}=1.5\omega_{0}$. For the chemical potentials we assumed
		$\mu_{\rm L}=1 \hbar \omega_{0}$ and $ \mu_{\rm R}=-5 \hbar \omega_{0}$. 
		The dimensionless coupling constant is chosen as $g=2.5$, and the internal bias voltage as
		$V_{\rm int} = 5 \hbar \omega_{0}$, wheras $\nu_{\rm L}=-\nu_{\rm R}= e V_{\rm int}/2$.
		}			
	\end{figure*}
	\section{Double quantum dot system (DQD)}
	The DQD that we treat in this section consits of two single dot levels coupled by a tunnel barrier.
	Again we assume a coupling to a single bosonic mode.
	The total Hamiltonian is composed of the oscillator part $H_{\rm osc}$, cf. eq. ~\eqref{eq.:H_osc}, 
	the electronic part $H_{\rm e}$ and an
	interaction part which describes the coupling between the oscillator and the two dots.  
	In contrast to the AHM, here the oscillator couples to the difference of 
	the occupation numbers with the coupling strength $\lambda$. The total Hamiltonian therefore reads
	\begin{equation}		\label{eq.:H_DQD}
		H_{\rm DQD} = H_{\rm e} + H_{\rm osc} 
		-\lambda x (\hat d^\dag_{\rm L} \hat d_{\rm L} -\hat d^\dag_{\rm R} \hat d_{\rm R}) ,
	\end{equation}
	containing the electronic part
	\begin{align}
	H_{\rm e} &= \sum_{k\alpha}\varepsilon_{k\alpha} \hat c^\dag_{k\alpha} \hat c_{k\alpha}
		+ \sum_{\alpha} \nu_\alpha  \hat d^\dag_\alpha \hat d_\alpha
		+ T_c   \hat d^\dag_{\rm L} \hat d_{\rm R} + T^*_c   \hat d^\dag_{\rm R} \hat d_{\rm L}  \nonumber \\
	&\hspace{1ex}+ \sum_{k\alpha}\big[ V_{k\alpha} \hat c^\dag_{k\alpha} \hat d_{\alpha}
		+ V^*_{k\alpha} \hat d^\dag_{\alpha} \hat c_{k\alpha}\big]. 
	\end{align}
	In this model, $\hat d^\dag_\alpha$/$\hat d_\alpha$ denotes the creation/annihilation operator of 
	the $\alpha$th ($\alpha\in\{\rm L, R\}$) dot. $\nu_\alpha$ designates the corresponding level energy and $T_c$ describes 
	the tunnel coupling matrix element between the two dots. Note that we include no Coulomb interaction terms here.

	The stationary average of the population difference $\ev{\sigma_z}$ corresponds to 
	the electronic force operator $\hat F$ of the generic model, eq. \eqref{eq.:H_gen}.
	The occupation of the $\alpha$th dot 
	\begin{align}
	\ev{\hat n_\alpha} &= -i\frac{1}{2\pi}\int d\omega\; G^<_{\alpha,\alpha}(\omega)
	\end{align}
	can be calculated by using Keldysh's equation \eqref{eq.:Keldysh}, see section~\ref{sec.:DQD_green}. 
	Therewith the occupation difference of the left/right dot follows from 
	\begin{eqnarray}
		\ev{\sigma_{z}} &=& \ev{n_{\rm L}} - \ev{n_{\rm R}} \nonumber\\ &=&
		\frac{\Gamma}{4\pi}  \Bigg[
		\int\limits_{-\infty}^{\mu_{\rm L}} d\omega \
		\frac{(\omega - \widetilde{\nu}_{\rm R})^{2} +  \left(\frac{\Gamma}{4} \right)^{2} - \abs{T_{C}}^{2}}{\omega^{4} + 2 A \omega^{2} + B^{2}}  \nonumber\\ && 
		\hspace{2ex} -
		\int\limits_{-\infty}^{\mu_{\rm R}} d\omega \
		\frac{(\omega - \widetilde{\nu}_{\rm L})^{2} +  \left(\frac{\Gamma}{4} \right)^{2} - \abs{T_{C}}^{2}}{\omega^{4} + 2 A \omega^{2} + B^{2}} \Bigg],   
	\end{eqnarray}
	with the abbreviations $\tilde\nu_{\rm L,R}=\nu_{\rm L,R}\mp\lambda x$ and
	\begin{align}
	A &= -\big[\abs{T_c}^2+\tilde\nu_{\rm L}^2-(\Gamma/4)^2\big], 
	\hspace{0.2cm} B= \big[\abs{T_c}^2+\tilde\nu_{\rm L}^2+(\Gamma/4)^2\big],
	\end{align}
	whereas we assumed $\widetilde{\nu}_{\rm R} = - \widetilde{\nu}_{\rm L}$.
	Calculating the integrals leads to
	\begin{eqnarray}
		\ev{\sigma_{z}} &=& \frac{\Gamma}{4\pi} \sum_{\alpha \in \rm L,R} \Bigg[ 
		2  \widetilde{\nu}_{\rm L} \mathcal I_2(\mu_{\rm \alpha},A,B) \nonumber \\ & &
		+ \sgn(\nu_{\alpha} \nu_{\rm L}) \bigg\{
		\mathcal I_3(\mu_{\rm \alpha},A,B) +   \nonumber \\ & &
		\left(\widetilde{\nu}_{\rm L}^{2} + \left(\Gamma/4 \right)^{2} - \abs{T_{C}}^{2}\right) 
		\mathcal I_1(\mu_{\rm \alpha},A,B) \bigg\}  \Bigg]. \nonumber\\
	\end{eqnarray}
	They are expressed in terms of the auxiliary functions
	$\mathcal I_j$
	defined in appendix~\ref{sec.:DQD_aux_int}.

	\subsection{Langevin equation}
	When applying the generic model, cf. eq.~\eqref{eq.:gen_model_langevin}, to the DQD Hamiltonian $H_{\rm DQD}$, 
	we obtain the Langevin equation
	\begin{align}
	m\ddot x_t -F_ {\rm eff}(x_t )+\dot x_t A[x](t) &= \xi_t \label{eq.:DQD_stoch_motion_eq}
	\end{align}
	with the effective force $F_{\rm eff}$ and the friction $A[x](t)$. In contrast to the AHM, 
	the effective force is affected by the population difference and not only by the occupation number. 
	Explicitly,
	\begin{align}
	F_ {\rm eff}(x_t ) &= -m\omega_0^2 x_t + \lambda \ev{\tilde\sigma_z(t)}, \\
	A[x](t) &= 2\lambda^2 \int_{t_0}^{t}dt'\;t'
		\Im\ev{\delta\tilde\sigma_z(t)\delta\tilde\sigma_z(t')}\label{eq.:def_DQD_friction}.
	\end{align}
	In appendix~\ref{sec.:DQD_friction} we derive the explicit expression for the friction.
	In the case of infinite bias (IB) and $\widetilde{\nu}_{\rm R} = - \widetilde{\nu}_{\rm L}$ we obtain
	\begin{equation}
		A_{\rm IB}[x](t) = - 8 \ \left|T_{c}\right|^{2} \frac{\lambda^{2}}{\Gamma} \ \widetilde{\nu}_{\rm L}  \
		\frac{\widetilde{\nu}_{\rm L}^{2} + \left|T_{c}\right|^{2} + 5 \frac{\Gamma^{2}}{16}}{\left(\widetilde{\nu}_{\rm L}^{2} 
		+ \left|T_{c}\right|^{2} 
		+  \frac{\Gamma^{2}}{16}\right)^{3}}.
	\end{equation}
	In contrast to the AHM the friction does not disappear for infinite bias.
	The second difference that we can recognise by regarding the prefactor $\widetilde{\nu}_{\rm L}$
	is that we obtain regions where the friction is {\em negative}. The latter also holds for the finite bias case.

	The real part of the correlation function according to the population difference $\sigma_{z}=d^\dag_{\rm L}d_{\rm L}-d^\dag_{\rm R}d_{\rm R}$ determines the correlation function 
	of the stochastic force
	\begin{align}
	\ev{\xi_{t}\xi_{t'}} = \lambda^2 2\Re\ev{\delta\tilde\sigma_z[x](t)\delta\tilde \sigma_z[x](t')}\label{eq.:DQD_stoch_forces_average},
	\end{align}
	with fluctuation $\delta\tilde\sigma_{z}[x](t)=\tilde\sigma_{z}[x](t)-\ev{\tilde\sigma_{z}[x](t)}$.
	\subsection{Fixed point analysis} \label{subsec.:DD_Fix}
	The effective potential determines the behavior of the oscillator trajectories in the phase space ($\ev{p}$--$\ev{x}$--plane), 
	as seen in section~\ref{subsec.:AHM_Eff_pot} for the Anderson Holstein model.
	In the following we examine the differential equation of the system by studying its fixed points \cite{Armen2006,Rodrigues2007_2}.
	Some further theoretical details of this kind of investigation are explained in appendix \ref{sec.:DQD_fix_point}.%

	For the double dot we obtain the dynamical system
	\begin{eqnarray}
		\ev{\dot{x_{t}}}  &=& \frac{1}{m} \ev{p_{t}} \nonumber \\
		\ev{\dot{p_{t}}}  &=&  \frac{\omega_{0}}{l_{0}}
		\left[-\frac{\ev{x_{t}}}{l_{0}} + g \ev{\sigma_z(t)} - 
		\frac{\ev{p_{t}}}{\omega_{0} l_{0}}  \ l_{0}^{2} A[\ev{x}](t) \right].\quad
	\end{eqnarray}
	Fixed points occur under the condition $\ev{\dot{p_{t}}} = \ev{\dot{x_{t}}}= 0$, i.e. $\ev{p_{t}}=0$ and 
	following from that, the fixed points position coordinates are equal to the roots of the effective force
	$$ F_{\rm eff}(\ev{x_{t}}) = -\frac{\ev{x_{t}}}{l_{0}} + g \ev{\sigma_z(t)} = 0.$$
	The Jacobian matrix is obtained from
	\begin{center}
		$J^{\ast} = \left(
		\begin{array}{*{2}{c}}
			0 &  1 \\
			\frac{\omega_{0}}{l_{0}}\left[-\frac{1}{l_{0}} + g \frac{\partial}{\partial \ev{x_{t}}} \ev{\sigma_z(t)}\big|_{\ev{x^{\ast}}}\right] & - A[\ev{x^{\ast}}] \\
		\end{array}
		\right)$
	\end{center}
	evaluated at the fixed point $\ev{x^{\ast}}$, whereby $\ev{p^{\ast}} = 0$.\\
	Determinant $\Delta$ and trace $\tau$ become
	\begin{equation}
		\Delta = \frac{\omega_{0}}{l_{0}}\left[\frac{1}{l_{0}} - 
		g \frac{\partial}{\partial \ev{x_{t}}} \ev{\sigma_z(t)}\big|_{\ev{x^{\ast}}}\right], \hspace{0.3cm}
		\tau =  - A[\ev{x^{\ast}}].
	\end{equation}
	The trace decides about the stability of a fixed point and is equal to the negative friction here.
	For the case without friction the trace is equal to zero, therefore only centers occur in the phase plane.
	In the case with friction the trace can be either positive or negative leading to both,
	stable and unstable fixed points. For comparison see Figure \ref{fig.:DQD_phase_space_comparison},
	where row A  depicts the results of a fixed point analysis.

	The effective force $F_{\rm eff}$ is plotted in the upper part of each plot in row A together with trace and determinant.
	Therewith the characteristics of the fixed points are determined and it is possible to predict the shape of the phase space portrait.  
	In the lower parts of the plots in row A these predictions are illustrated, 
	fixed points are marked by black lines.
	For small tunnel coupling $T_{c}$ (diagram A1) we obtain seven fixed points.
	These can be characterised as three stable and one unstable spirals, each seperated by one of the three saddle points.
	Stable spirals correspond to the different rest positions for the oscillator 
	and the saddle points to the points of instable balance. 
	Increasing the tunnel coupling leads to five fixed points. 
	In graph A2 we obtain a stable spiral in the middle
	enclosed by two saddle points and followed by a stable spiral on each side. 
	In the diagrams A3 and A4 the mean point changes to an unstable spiral.
	With further increased tunnel coupling (A3) there remain only three fixed points and
	for even higher values of $T_{c}$ (A4) the number of zeros in the effective force reduces to one.
	For the oscillator the latter means that it is shifted to a new rest position independent from its inital position.
	\subsection{Phase space portraits}
	In Figure \ref{fig.:DQD_phase_space_comparison} the rows B and C depict phase space portraits 
	for the double dot system without and with friction. 
	The four columns correspond to four different values of the tunnel coupling $T_{c}$. As initial condition, the momentum was set to zero 
	and the positions were chosen in order to show the various shapes of the trajectories.
	In the case without friction we recognize periodic cycles which are stable and run around one or more fixed points. 
	These centers were also expected from the analysis in section \ref{subsec.:DD_Fix}. 
	The fixed points correspond to certain states of the double dot system. If the left dot is occupied the rest position
	of the oscillator is shifted to the right and correspondingly to the left for an occupied right dot. 
	These points turn to stable spirals when the friction is turned on. 
	The states when both dots are occupied or empty correspond to the fixed point in the middle.

	The lowermost row shows what happens when we include the friction in our calculations. In contrast to the single dot system, here
	the friction has positive as well as negative values depending on the position of the oscillator.
	This means that the oscillator is either decelerated or accelerated. Both can be 
	interpreted as inelastic jumps of the electrons, where energy is transfered between the electrons and the oscillator
	in both directions like it has been observed in \cite{Rodrigues2007}. There, the authors consider a
	resonator coupled to a superconducting single electron transistor (SSET). 
	As a result of the interplay of positive and negative damping in a certain 
	parameter range they observed limit cycles and bistability in the phase plane, 
	in our work we obtain a likewise behaviour for the oscillator.
	In contrast to our semiclassical approximation they investigate the Wigner function of the system
	with numerical master equations.
	They compare these results to a mean field evaluation
	of the expectation value of the oscillator position \cite{Rodrigues2007_2}.
	For weak coupling the mean field approach gives
	quantitatively correct results, and for higher coupling it
	still describes the dynamics qualitatively correct.
	These results suggest that our use of average oscillator positions and momenta 
	is qualitatively correct for a description of the oscillator instabilities.

	Consider again row C in Figure \ref{fig.:DQD_phase_space_comparison}, 
	where the results for the dynamics of the oscillator with friction are plotted. 
	The outer left and right stable spiral do not change by increasing the tunnel coupling $T_{c}$. 
	By contrast the oscillator's behaviour between these stable rest positions changes a lot.
	In graph C1 we observe a stable and an unstable spiral, like we expect from the fixed point analysis. 
	In the neighbourhood of the unstable fixed point the friction is negative, so 
	the oscillator trajectory is repelled and ends up in the right stable spiral.
	In diagram C2 we recognize that the latter path becomes stable. The limit cycle appears when  
	the unstable spiral has disappeared and exists as long as the tunneling coupling is in the range of 
	$0.42 \leq \abs{T_{c}}^2 / \omega_{0}^{2} \leq 0.5$. 
	There we observe a bistability:
	as the initial position gets closer to the fixed point the limit cycle
	turns into a stable spiral.
	By further increasing $T_{c}$, the middle spiral becomes unstable and a second limit cycle appears (C3).  
	This limit cycle with a smaller radius exists in the range of $0.49 \leq \abs{T_{c}}^2 / \omega_{0}^{2} \leq 0.54$. 
	For $\abs{T_{c}} / \omega_{0}^{2} \simeq 0.49$ the system undergoes a Hopf bifurcation \cite{Strogatz2000}, 
	which happens when a pair of complex eigenvalues from the dynamical system, which determine the evolution in the phase plane, see section \ref{sec.:DQD_fix_point},
	cross the imaginary axis from the left to the right half-plane. 
	In other words, the trace $\tau$ changes its sign and at the bifurcation point the eigenvalues are purely 
	imaginary $\lambda_{1,2} = \mp i 2 \sqrt{\Delta}$, see equation \eqref{eq.:eigenvalues}.
	In the last graph of row C both limit cycles have disappeared and the oscillators path ends up in the 
	left or right stable point.

	In our calculations we choose $\Gamma = 3 \omega_{0}$, standing in some contrast to our adiabatic approach, 
	which implies a slow oscillator ($\Gamma \gg \omega_{0}$). 
	If $\Gamma \sim \omega_{0}$ the interaction between the current and the oscillator is strongest \cite{Rodrigues2005}, 
	because both act on the same timescale.
	Interesting effects still appear with a slightly enlarged $\Gamma$
	like in our plots, 
	but for $\Gamma \gg \omega_{0}$ there remains only one stable spiral. This means, that
	the oscillator rest position is shifted from its ground position caused by the stochastic processes initiated by the current.
	We presume that our approach is useful also for a comparative fast oscillator and 
	we will accomplish further investigations with a non-adiabatic approximation to reconsider our results.

	\section{Summary}
	We have derived a stochastic equation of motion that describes the dynamics of a single oscillator coupled to an electronic environment out of equilibrium. We studied two cases, namely the single dot level and double dot (two--level) electronic system.
	For both cases we have explained the features of effective potential and friction for the ensemble averaged oscillator motion. The effects we recognize fit well together with former works. In the DQD model limit cycles and bistabilities appear.
	
	Until now the master equation has been used for most investigations of the oscillator behaviour in NEMS. We have used a method that gives us access to regions where the master equation has problems: we naturally include finite bias, and arbitrary electron coupling to external reservoirs.
	
	We had to stay in a regime with a relatively fast oscillator in order not to miss the interesting physical effects. The validity of our method in this regime could still be improved with a non-adiabatic calculation.

	\section{Acknowledgements}
	This work was supported by project DFG BR 1528/5-2, the  WE Heraeus foundation and the Rosa Luxemburg foundation.
	We acknowledge discussions with C. Emary.\appendix
	\section{\label{sec.:gen_mod}Path integral representation of the reduced density matrix}
	The reduced density matrix $\rho(t)=\tr_{\rm B}\chi(t)$, obtained from the total density matrix $\chi(t)$ by tracing out the bath degrees of freedom, describes the mechanic subsystem $\mathcal{H}_{\rm osc}$, cf.~\eqref{eq.:H_gen}. By using a factorising initial condition $\chi(t_0)=\rho(t_0)\otimes\rho_{\rm B}$ the elements of the reduced density matrix can be expressed in a path integral representation\cite{Schulman2005,Weiss2008}
	\begin{align}
	\braketop{q}{\rho_{\rm osc}(t)}{q'}
	&=\int dq_0\;dq'_0\;\braketop{q_0}{\rho_{\rm osc}(t_0)}{q'_0}\nonumber\\
		\times \int_{q(t_0)}^{q(t)} 
	&\mathcal{D}q(\tau)\int_{q'(t_0)}^{q'(t)} \mathcal{D}^*q'(\tau)\;
		e^{i(S_q-S_{q'})}\mathcal{F}[q,q'](\tau)\label{eq.:red_dense_matrix_path_rep}.
	\end{align}
	Hereby $S_q = \int_{t_0}^t dt'\; \big[\frac12 m \dot q_{t'}^2-V_{\rm osc}(q_{t'})\big]$ denotes the classical action in the path integral for~eq.~\eqref{eq.:red_dense_matrix_path_rep} and the Feynman--Vernon influence functional is defined by
	\begin{align}
	\mathcal{F}[q,q'](t_0,t)=\tr_{\rm B}\big\{\hat U^\dag[q'](t,t_0)\hat U[q](t,t_0)\rho_{\rm B}\big\}.
	\end{align}
	The time--evolution operators  are defined by
	\begin{align}
	\hat U[q](t,t_0)=\Tfor \exp\big[-i\int_{t_0}^{t}dt'\; \mathcal{H}_{\rm res}[q](t')\big]
	\end{align}
	corresponding to the reservoirs $\mathcal{H}_{\rm res}[q](t)=\mathcal{H}_{\rm e}-\hat F q_t$.
	In the next step we establish an effective interaction picture by
	\begin{align}
	\tilde U[q](t,t_0)
	&= e^{i\mathcal{H}_0(t-t_0)} \; \hat U[q](t,t_0)\nonumber\\
	&= \Tfor \exp\big[-i\int_{t_0}^{t}dt'\;\tilde V[q](t')\big],\nonumber\\
	\tilde V[q](t)&=
		e^{i\mathcal{H}_0(t-t_0)} 
		V[q](t) e^{-i\mathcal{H}_0(t-t_0)} \label{eq.:gen_time_evol}.
	\end{align}
	Hereby the reservoir is decomposed in an unperturbed part $\mathcal{H}_0 = \mathcal{H}_{\rm e}-\hat F x_0$ and the perturbation $V[q](t)= -\hat F(t{\dot x}_0+\frac12 y_t)$ and $V[q'](t)= -\hat F(t{\dot x}_0-\frac12 y_t)$, where  we wrote $x_t \approx x_0+t\dot x_0$ using the adiabatic approximation \eqref{eq.:gen_adiabatic_approx}.
	
	By using eq.~\eqref{eq.:gen_time_evol}, the influence functional can be expanded to second order in perturbation theory. Denoting $\ev{\;\cdot\;}=\tr_{\rm B}\{\rho_{\rm B}\;\cdot\;\}$, $\tilde V_1'\rdef \tilde V[q'](t_1)$ and $\tilde V_1\rdef \tilde V[q](t_1)$, the Feynman--Vernon influence functional reads
	\begin{align}
	&\mathcal{F}[q,q'](t_0,t) = \ev{\tilde U[q'](t,t_0)\tilde U[q](t,t_0)}\nonumber\\[1ex]
	&= \ev{
		\big[\Tback e^{i\int_{t_0}^{t}dt_1\;\tilde V'_1}\big]
		\big[\Tfor e^{-i\int_{t_0}^{t}dt_1\;\tilde V_1}\big]
		}\nonumber\\
	&= 1 + i\int_{t_0}^{t}dt_1\;\ev{\tilde V'_1-\tilde V_1}\nonumber\\
	&\hspace{2ex}+\int_{t_0}^{t}dt_1\int_{t_0}^{t_1}dt_2\;
		\ev{(\tilde V'_1-\tilde V_1)\tilde V_2-\tilde V'_2(\tilde V'_1-\tilde V_1)}.
	\end{align} 
	When plugging in the definition of the perturbation (interaction picture) and taking into account that the force operator $\hat F$ is hermitian, we obtain  
	\begin{align}
	&\mathcal{F}[x+y/2,x-y/2](t_0,t)
		= 1 + i\int_{t_0}^{t}dt_1\;y_{t_1}\nonumber\\
	&\hspace{2ex}\times
		\bigg[\ev{\tilde F(t_1)} -2\dot x_0 \int_{t_0}^{t_1}dt_2\;t_2
		\Im\ev{\tilde F(t_1)\tilde F(t_2)}\bigg]\nonumber\\
	&\hspace{2ex}-\int_{t_0}^{t}dt_1\int_{t_0}^{t_1}dt_2\;y_{t_1}
		\Re\ev{\tilde F(t_1)\tilde F(t_2)}y_{t_2} \label{eq.:gen_influece_functional}.
	\end{align}
	Performing a cluster expansion \cite{VanKampen2008}, the influence functional can be expressed in terms of the influence phase
	\begin{align}
	&\Phi[x,y](t_0,t)\equiv\nonumber\\
	&\hspace{2ex}-\ln\mathcal{F}[x+y/2,x-y/2](t_0,t)
		= - i\int_{t_0}^{t}dt_1\;y_{t_1}\nonumber\\
	&\hspace{2ex}\times
		\bigg[\ev{\tilde F(t_1)} -2\dot x_0 \int_{t_0}^{t_1}dt_2\;t_2
		\Im\ev{\delta\tilde F(t_1)\delta\tilde F(t_2)}\bigg]\nonumber\\
	&\hspace{2ex}+\int_{t_0}^{t}dt_1\int_{t_0}^{t_1}dt_2\;y_{t_1}
		\Re\ev{\delta\tilde F(t_1)\delta\tilde F(t_2)}y_{t_2},
	\end{align}
	whereby $\delta\tilde F(t)=\tilde F(t)-\ev{\tilde F(t)}$ denotes the fluctuation around $\ev{\tilde F(t)}$.
	This equation can be easily verified by expanding the exponential $\exp[-\Phi[x,y](t_0,t)]$ to second order in perturbation theory and comparing with eq.~\eqref{eq.:gen_influece_functional}. 
	
	Together with the classical action, we define an effective action functional by
	\begin{align}
 	&\mathcal{A}[x,y] (t_0,t)
		\rdef S_{x+y/2} - S_{x-y/2} + i\Phi[x,y](t_0,t)\nonumber\\
	&= \int_{t_0}^t dt_1\; \big[
		m \dot x_1 \dot y_1 -V_{\rm osc}(x_1+\frac{y_1}{2})+V_{\rm osc}(x_1-\frac{y_1}{2})
	\big]\nonumber\\
	&\hspace{2ex}+ i\Phi[x,y](t_0,t),
	\end{align}
	whereby the reduced density matrix expressed in terms of the new variables, eq. \eqref{eq.:gen_xy_expansion}, reads
	\begin{align}
	&\braketop{x+\frac{y}{2}}{\rho(t)}{x-\frac{y}{2}}
		=\int dx_0\;dy_0\;\braketop{x_0+\frac{y_0}{2}}{\rho(t_0)}{x_0-\frac{y_0}{2}}\nonumber\\
	&\hspace{2ex}\times \int_{x(t_0)}^{x(t)}\mathcal{D}x(\tau) \int_{y(t_0)}^{y(t)}\mathcal{D}^*y(\tau)\;
		e^{i\mathcal{A}[x,y] (t_0,t)}.
	\end{align}
	In a semiclassical approach we assume small deviations of the off--diagonal trajectories $y$ from the diagonal ones. Thus the potential difference leads in second order in $y$ to 
	\begin{align}
	V_{\rm osc}(x_1+\frac{y_1}{2})-V_{\rm osc}(x_1-\frac{y_1}{2})
		= V'_{\rm osc}(x) + \mathcal{O}(y^3).
	\end{align}
	This approximation is exact for quadratic potentials, as we treat in this paper. Furthermore, with the boundary conditions $y(t_0)=y(t)=0$ and integration by parts the effective action functional is quadratic in $y$,
	\begin{align}
 	&\mathcal{A}[x,y] (t_0,t)
		= -\int_{t_0}^t dt_1\;y_1 \bigg[
			m \ddot x_1  +V'_{\rm osc}(x_1) -\ev{\tilde F(t_1)}
	\nonumber\\
	&\hspace{2ex}+2\dot x_0 \int_{t_0}^{t_1}dt_2\;t_2\Im\ev{\delta\tilde F(t_1)\delta\tilde F(t_2)}
		\bigg]\nonumber\\
	&\hspace{2ex}+ i\int_{t_0}^{t}dt_1\int_{t_0}^{t_1}dt_2\;y_1
		\Re\ev{\delta\tilde F(t_1)\delta\tilde F(t_2)}y_2]\nonumber\\
	&\equiv -\int_{t_0}^t dt_1\;y_1 K_1[x]
		+\frac{i}{2}\int_{t_0}^{t}dt_1\int_{t_0}^{t_1}dt_2\;y_1 L_{1,2}[x]y_2 \label{eq.:gen_action_functional}.
	\end{align}
	Completing the square of the integral kernel \cite{Weiss2008}
	$\int D^*y \exp\{i\mathcal{A}[x,y]\}$, the resulting path integral describes a stochastic process with Langevin equation
	\begin{align}
	 K_t[x] &=
		m \ddot x_t  +V'_{\rm osc}(x_t) -\ev{\tilde F(t)}\nonumber\\
	&\hspace{2ex} +2\dot x_0 \int_{t_0}^{t}dt'\;t'\;\Im\ev{\delta\tilde F(t)\delta\tilde F(t')} = \xi_t 
	\end{align}
	with $\xi_t$ a Gaussian stochastic force. To obtain a selfconsistent equation of motion, we finally replace $x_0$
	by $x_t$. This substitution concerns also the unperturbed Hamiltonian which leads to $\mathcal H_0\rightarrow \mathcal H_{\rm e} -\hat F x_t$.
	The Langevin equation then reads 
	\begin{align}
	m \ddot x_t  &+V'_{\rm osc}(x_t) -\ev{\tilde F(t)}\nonumber\\
	&+2\dot x_t \int_{t_0}^{t}dt'\;t'\;\Im\ev{\delta\tilde F(t)\delta\tilde F(t')}= \xi_t .
	\end{align}
	The term quadratic in $y$ in eq.~\eqref{eq.:gen_action_functional} determines the correlation function of the stochastic force,
	\begin{align}
	\ev{\xi_t\xi_{t'}} &=
		2\Re\ev{\delta\tilde F(t)\delta\tilde F(t')}.
	\end{align}
	
	\section{\label{sec.:AHM_Corr} Green's functions of the single dot}
	The single dot lesser Green's function without coupling in energy space is\cite{Haug2008}
	\begin{align}
	G^<(\omega)  = 
		i\frac{\Gamma_{\rm L}f_{\rm L}(\omega)+\Gamma_{\rm R}f_{\rm R}(\omega)}
		{(\omega-\varepsilon_x)^2+\Gamma^2/4}.
	\end{align}
	Here we have used the adiabatic approximation for
	the center of mass coordinate, thus the coupling
	to the oscillator simply shifts the level $\varepsilon_{\rm d}$
	by $\lambda\ev{x}$, so we have to replace $\varepsilon_{\rm d}$ by the shifted energy $\varepsilon_x = \varepsilon_{\rm d} -\lambda\ev{x}$.
	The greater Green's function is correspondingly
	\begin{align}
	G^>(\omega)  &=  -i\frac
		{\Gamma_{\rm L}[1-f_{\rm L}(\omega)]+\Gamma_{\rm R}[1-f_{\rm R}(\omega)]}
		{(\omega-\tilde\varepsilon)^2+\Gamma^2/4}\nonumber\\
	&=G^<(\omega) 
		-i\frac{\Gamma}{(\omega-\tilde\varepsilon)^2+\Gamma^2/4}.
	\end{align}
	Hereby $f_\alpha(\omega)=f(\omega-\mu_\alpha)=1/[\exp\{\beta(\omega-\mu_\alpha)\}+1]$ denote Fermi functions with lead index $\alpha$, inverse temperature $\beta$ and chemical potential $\mu_\alpha$.
	The time-dependent Green's functions are obtained by Fourier transformation as
	$G^{\lessgtr}(t) = \int\frac{d\omega}{2\pi}e^{-i\omega t}G^\lessgtr(\omega)$.
	In the zero--temperature limit we have to replace
	$f_\alpha(\omega) = \Theta(\mu_\alpha-\omega)$.
	Then for $t\neq 0$ an integration leads to
	\begin{align}
	&G^\lessgtr(t)= 
		\frac{e^{-i\tilde\varepsilon t}}{2\pi}\sum_{\alpha}\frac{\Gamma_\alpha}{\Gamma} \bigg[\nonumber\\
	&\hspace{2ex}-e^{+\Gamma/2\abs{t}}\Eone\big\{[i\Omega_\alpha\sgn(t)+\Gamma/2]\abs{t}\big\}\sgn(t)\nonumber\\
	&\hspace{2ex}+e^{-\Gamma/2\abs{t}}\Eone\big\{[i\Omega_\alpha\sgn(t)-\Gamma/2]\abs{t}\big\}\sgn(t)\nonumber\\
	&\hspace{2ex}\pm 2\pi ie^{-\Gamma/2\abs{t}}\Theta(\pm\Omega_\alpha)
		\bigg]\bigg|_{\Omega_\alpha=\mu_\alpha-\varepsilon_x},
	\end{align}
	where the first exponential integral is defined by
	\begin{align}
	\Eone(x) &=
		\int_1^\infty dt\; \frac{e^{-xt}}{t}.
	\end{align}
	For the case $t=0$ the lesser Greens function reads
	\begin{align}
	-iG^<(t=0)&= 
		\frac{1}{2\pi}\int_{-\infty}^\infty d\omega\;
		\sum_{\alpha}\frac{\Gamma_\alpha \Theta(\mu_\alpha -\omega)}{(\omega-\varepsilon)^2+\Gamma^2/4} \nonumber\\
	&=\frac{1}{2}-\frac{1}{\pi}\sum_{\alpha}\frac{\Gamma_\alpha}{\Gamma}\arctan\big[\frac{2}{\Gamma}(\tilde\varepsilon-\mu_\alpha)\big].
	\end{align}
	
	For finite temperature one has to regard Fermi functions instead of the Heaviside theta function. 
	By virtue of the residue theorem one finds
	\begin{align}
	-i&G^<(t=0)\nonumber\\
	&= \frac12 -\frac{1}{\pi}\sum_\alpha\frac{\Gamma_\alpha}{\Gamma}
		\Im\Psi\bigg(\frac12 +\frac{\beta\Gamma}{4\pi} 
		+i\frac{\beta(\tilde\varepsilon-\mu_\alpha)}{2\pi}\bigg),
	\end{align}
	with the digamma function $\Psi$.
	\section{\label{sec.:DQD_AppX} Characteristics and calculations for the double dot}
	\subsection{\label{sec.:DQD_green} Green's function of the DQD}
	The Green's functions in the frequency domain according to the Hamiltonian in equation~(\ref{eq.:H_DQD}) 
	is derived via the equation of motion method. There the Green's function $G$ is defined as resolvent 
	of the Hamiltonian $H_{0}$ via
	\begin{equation}
		(\omega \id - H_{0}) G(\omega) = \id
	\end{equation}
	Denoting the bath states with $|\phi_{\lambda}\rangle  = |\lambda\rangle$ the previous equation yields
	\begin{equation}
		\langle\lambda|(\omega\id - H_{0})G(\omega)|\lambda'\rangle = \delta_{\lambda,\lambda'}
	\end{equation}
	Taking the matrix elements and inserting the Hamiltonian leads to a set of equations, 
	from whom the Green's functions are derived. \\
	Here, the electron-phonon coupling is described adiabatically, 
	so the interaction part $H_{SB}$ came in by shifting the dot level energies $\nu_{\rm{L,R}} \rightarrow \widetilde{\nu}_{\rm{L,R}}$
	with $\widetilde{\nu}_{\rm{L,R}} = \nu_{\rm{L,R}} \mp \lambda x$. Finally the dot Green's function is given through
	$$\textbf{G}_{\rm{D}}(\omega) = 
	\begin{pmatrix}
		G_{\rm{LL}}(\omega) & G_{\rm{LR}}(\omega) \\ G_{\rm{RL}}(\omega) & G_{\rm{RR}}(\omega) \\  
	\end{pmatrix}
	$$
	with the elements
	\begin{eqnarray}		\label{eq.:ddgreenfunctions}
		G_{\rm{LL}}(\omega)& = & \frac{\omega - \widetilde{\nu}_{\rm{R}} - \Sigma_{\rm{R}}(\omega)}{\left[\omega - \widetilde{\nu}_{\rm{L}} - \Sigma_{\rm{L}}(\omega)\right] \left[\omega - \widetilde{\nu}_{\rm{R}} - \Sigma_{\rm{R}}(\omega)\right] - \left|T_{c}\right|^{2}} \nonumber \\
		G_{\rm{RR}}(\omega)& = & \frac{\omega - \widetilde{\nu}_{\rm{L}} - \Sigma_{\rm{L}}(\omega)}{\left[\omega - \widetilde{\nu}_{\rm{L}} - \Sigma_{\rm{L}}(\omega)\right] \left[\omega - \widetilde{\nu}_{\rm{R}} - \Sigma_{\rm{R}}(\omega)\right] - \left|T_{c}\right|^{2}} \nonumber \\
		G_{\rm{LR}}(\omega)& = & \frac{T_{c}}{\omega - \widetilde{\nu}_{\rm{L}} - \Sigma_{\rm{L}}(\omega)} \ G_{\rm{RR}}(\omega) \nonumber \\
		G_{\rm{RL}}(\omega)& = & \frac{T^{\ast}_{c}}{\omega - \widetilde{\nu}_{\rm{R}} - \Sigma_{\rm{R}}(\omega)} \ G_{\rm{LL}}(\omega)
	\end{eqnarray}
	Hereby we introduced the self energy $\Sigma_{\alpha} , (\alpha \in \rm{R,L})$ corresponding to the left or the right dot with
	\begin{eqnarray}
		\Sigma_{\alpha} (\omega) = \sum_{k} \abs{ V_{k\alpha} }^{2} \ g_{k\alpha, k\alpha}(\omega).
	\end{eqnarray}
	$g_{k\alpha, k\alpha}$ is the undisturbed Green's function for the leads. 
	We derive the associated retarded and the advanced Green's function by using the continuation rules
	$$G(\omega \pm i 0^{+}) \rightarrow G^{R,A}(\omega)$$
	The derivation of the lesser/greater Green's function is taking usage of the Keldysh equation:
	\begin{equation}		\label{eq.:Keldysh}
		G^{\lessgtr}_{\alpha \beta} = \sum_{\gamma} \ G^{R}_{\alpha,\gamma}(\omega) \ \Sigma^{\lessgtr}_{\gamma}(\omega) \ G^{A}_{\gamma, \beta}(\omega)
	\end{equation}
	whereas we assume both dots initially unoccupied. The lesser/greater self energy follows from
	\begin{eqnarray}
		\Sigma^{\lessgtr}_{\gamma}(\omega) & = & \sum_{k} \abs{ V_{k\gamma} }^{2} \ g^{\lessgtr}_{k\gamma, k\gamma}(\omega), \nonumber \\
		\Rightarrow \Sigma^{<}_{\gamma}(\omega) & = & i \Gamma_{\gamma} (\omega) \ f_\gamma(\omega), \nonumber \\
		\Sigma^{>}_{\gamma}(\omega) & = &- i \Gamma_{\gamma} (\omega) \ \left[1 - f_\gamma(\omega)\right].
	\end{eqnarray}
	
	\newpage
	\subsection{\label{sec.:DQD_friction} Calculation of the friction for the DQD}
	The friction is determined by the imaginary part of the correlation function of the double dot. 
	From equation \eqref{eq.:def_DQD_friction}, we start with expressing 
	the correlation function through Green's functions and use their Fourier transforms:
	\begin{align} \label{eq.:frictionDDstart}
	& A[x](t) \nonumber \\ & =  2 \lambda^{2} \ \sum_{\alpha,\beta} \ \left[2 \delta_{\alpha,\beta } -1\right] \ 
				\int dt' \ t' \ \frac{1}{2i} \ \ * \nonumber \\ & \hspace{0.5cm}
				\left[ G_{\beta \alpha}^{<}(t'-t)  G_{\alpha \beta}^{>}(t-t')  
					- G_{\beta \alpha}^{<}(t-t')  G_{\alpha \beta}^{>}(t'-t) \right] \nonumber \\ &=
			\frac{\lambda^{2}}{\pi} \Im \sum_{\alpha,\beta}  \left[2 \delta_{\alpha,\beta } -1\right] 
				\int d\omega_{1} \int d\omega_{2}  G_{\beta \alpha}^{<}(\omega_{1}) * \nonumber \\ & \hspace{0.5cm} G_{\alpha \beta}^{>}(\omega_{2}) 
				e^{i (\omega_{1} - \omega_{2}) t}  (-i)  \frac{\partial}{\partial \omega_{2}} 
											\int \frac{dt'}{2\pi}  e^{-i(\omega_{1}-\omega_{2}) t'} \nonumber \\ &=    
			\frac{\lambda^{2}}{\pi}  \sum_{\alpha,\beta}  
			\left[2 \delta_{\alpha,\beta} -1\right]  \int d\omega G^{<}_{\beta \alpha}(\omega)  \frac{\partial}{\partial \omega}  
				G^{>}_{\alpha \beta}(\omega).
	\end{align}
	In the last step a term was identified as the derivative of Dirac's delta, this result agrees with the solution 
	in the work by Mozyrsky et al.\cite{Mozyrsky2006}.\\
	The lesser/greater Green's function Keldysh equation follow from the Keldysh equation \eqref{eq.:Keldysh} and 
	we consider the tunneling rates to be frequency independent, therefore the self energies are ($T=0$):
	\begin{eqnarray}
		\Sigma^{<}_{\gamma}(\omega)&=&   i \Gamma_{\gamma} \   \Theta(\mu_{\gamma} - \omega), \nonumber \\
		\Sigma^{>}_{\gamma}(\omega)&=& - i \Gamma_{\gamma}  \  \Theta(\omega - \mu_{\gamma}). 
	\end{eqnarray}
	Because of the lengthy calculation, we just outline the calculation for the LL-term $\mathcal{T}_{\rm{LL}}$. 
	The other terms ($\mathcal{T}_{\rm{RL}},\mathcal{T}_{\rm{LR}} \ \mbox{and} \ \mathcal{T}_{\rm{RR}}$) can be derived in a similar way.
	Here, the product of retarded and advanced Green's functions equates the squared modulus of $G_{\alpha,\beta}$, so we can write
	\begin{align}
	G^{R}_{\rm{LL}}(\omega) \ G^{A}_{\rm{LL}}(\omega)  & = \abs{G^{R}_{\rm{LL}}(\omega)}^{2} = \abs{G^{A}_{\rm{LL}}(\omega)}^{2}, \nonumber \\
	G^{R}_{\rm{LR}}(\omega) \ G^{A}_{\rm{RL}}(\omega)  & = \abs{G^{R}_{\rm{LR}}(\omega)}^{2} = \abs{G^{A}_{\rm{RL}}(\omega)}^{2}.
	\end{align}
	For simplicity in the following we omit the superscript of the retarded Greens function and abbreviate the moduli by
	$$ \abs{G_{\rm{LL}}(\omega)}^{2} \equiv \abs{G^{R}_{\rm{LL}}(\omega)}^{2},\quad \abs{G_{\rm{LR}}(\omega)}^{2} \equiv \abs{G^{R}_{\rm{LR}}(\omega)}^{2}. $$
	We obtain for the first term of the friction 
	\begin{widetext}
		\begin{eqnarray}
			\mathcal{T}_{\rm{LL}}  &= &  \int d\omega 
			\bigg[ G^{<}_{\rm{LL}}(\omega)  \frac{\partial}{\partial \omega}  G^{>}_{\rm{LL}}(\omega)\bigg]  \nonumber \\ &=& 
			\frac{\Gamma^{2}}{4}      \int d\omega 
			\bigg[ \abs{G_{\rm{LL}}(\omega)}^{2}  \Theta(\mu_{\rm{L}} - \omega) 
			+ \abs{G_{\rm{LR}}(\omega)}^{2}  \Theta(\mu_{\rm{R}} - \omega) \bigg] * \nonumber \\ & & 
			\Bigg[ \left[ \frac{\partial}{\partial \omega}  \abs{G_{\rm{LL}}(\omega)}^{2} \right]  \Theta(\omega - \mu_{\rm{L}} )
			+ \abs{G_{\rm{LL}}(\omega)}^2  \delta(\omega - \mu_{\rm{L}} ) + 
			\left[\frac{\partial}{\partial \omega} \abs{G_{\rm{LR}}(\omega)}^{2} \right]  \Theta(\omega - \mu_{\rm{R}}) 
			\abs{G_{\rm{LR}}(\omega)}^{2}  \delta(\omega - \mu_{\rm{R}}) \Bigg]. \nonumber \\
		\end{eqnarray}
	\end{widetext}
	Some parts of the integration terms can directly be evaluated with the help of the delta and Heavyside functions. 
	By choosing the condition $\mu_{\rm{L}} > \mu_{\rm{R}}$, we get rid of a case distinction, which would be necessary for 
	two product terms which included a delta-function. After performing also an integration by parts,
	we arrive at 
	\begin{eqnarray}
		\mathcal{T}_{\rm{LL}} &=& \frac{\Gamma^{2}}{8}  
		\Bigg[            \left| G_{\rm{LL}}(\mu_{\rm{L}}) \right|^{4}  + \left| G_{\rm{LR}}(\mu_{\rm{R}}) \right|^{4} - \nonumber \\ & &
		\hspace{-0.9cm}   2 \left| G_{\rm{LL}}(\mu_{\rm{L}}) \right|^{2}    \left| G_{\rm{LR}}(\mu_{\rm{L}}) \right|^{2} + 
		4 \left| G_{\rm{LL}}(\mu_{\rm{R}}) \right|^{2}    \left| G_{\rm{LR}}(\mu_{\rm{R}}) \right|^{2} + \nonumber \\ & &
		\hspace{0.7cm}     \ 4 \int\limits_{\mu_{\rm{R}}}^{\mu_{\rm{L}}} d\omega  \left| G_{\rm{LL}}(\omega) \right|^{2}  
		\left[ \frac{\partial}{\partial \omega} \left| G_{\rm{LR}}(\omega) \right|^{2} \right]   \Bigg].
	\end{eqnarray}
	In the following the Green's functions, derived in section~\ref{sec.:DQD_green}, were inserted, whereas  we
	assume equal tunneling rates for the left and the right side, $\Gamma_{L} = \Gamma_{R} = \frac{1}{2}\Gamma$. 
	Then a number of integrations by parts is performed to dispose of the derivation in the integral term. 
	So we obtain a closed expression, whereas $N(\omega)$ abbreviates the denominator of the Green's function
	\begin{eqnarray}
		\mathcal{T}_{\rm{LL}} &=&
		\frac{\Gamma^{2}}{8} \ \Bigg[ \frac{\left[(\mu_{\rm{L}} - \widetilde{\nu}_{\rm{R}})^{2} + 
		\frac{\Gamma^{2}}{16}\right]^{2}}{N(\mu_{\rm{L}})^{2}} + \frac{\left| T_{c} \right|^{4}}{N(\mu_{\rm{R}})^{2}} + \nonumber \\ & & \hspace{-0.5cm}
		2 \left| T_{c} \right|^{2}  \frac{\left[(\mu_{\rm{R}} - \widetilde{\nu}_{\rm{R}})^{2} + \frac{\Gamma^{2}}{16}\right]}{N(\mu_{\rm{R}})^{2}}
		- 4  \left| T_{c} \right|^{2}  \int\limits_{\mu_{\rm{R}}}^{\mu_{\rm{L}}} d\omega  
		\frac{(\omega - \widetilde{\nu}_{\rm{R}})}{N(\omega)^{2}} \Bigg]. \nonumber \\
	\end{eqnarray}
	In an analogue way the other terms in \eqref{eq.:frictionDDstart} can be derived, and finally the solution for the friction is
	\begin{widetext}
		\begin{eqnarray}			\label{eq.:frictionDD}
			A[x](t)& = & 
			\frac{\lambda^{2}}{\pi}  \frac{\Gamma^{2}}{8} 
			\Bigg[ \frac{\left[(\mu_{\rm{L}} - \widetilde{\nu}_{\rm{R}})^{2} + \frac{\Gamma^{2}}{16}\right]^{2}}{N(\mu_{\rm{L}})^{2}} 
			+ \frac{\left[(\mu_{\rm{R}} - \widetilde{\nu}_{\rm{L}})^{2} + \frac{\Gamma^{2}}{16}\right]^{2}}{N(\mu_{\rm{R}})^{2}} 
			+ \frac{\left| T_{c} \right|^{4}}{N(\mu_{\rm{L}})^{2}} + \frac{\left| T_{c} \right|^{4}}{N(\mu_{\rm{R}})^{2}} 
			+ 2 \left| T_{c} \right|^{2}  \frac{\left[(\mu_{\rm{R}} - \widetilde{\nu}_{\rm{R}})^{2} + \frac{\Gamma^{2}}{16}\right]}{N(\mu_{\rm{R}})^{2}} 
			\nonumber \\ & & 
			- 2 \left| T_{c} \right|^{2}  \frac{\left[(\mu_{\rm{L}} - \widetilde{\nu}_{\rm{R}})^{2} + \frac{\Gamma^{2}}{16}\right]}{N(\mu_{\rm{L}})^{2}} 
			- 4 \left| T_{c} \right|^{2}  \frac{\left[(\mu_{\rm{R}} - \widetilde{\nu}_{\rm{R}})  
			(\mu_{\rm{R}} -\widetilde{\nu}_{\rm{L}}) - \frac{\Gamma^{2}}{16}\right]}{N(\mu_{\rm{R}})^{2}}    
			- 8 \left| T_{c} \right|^{2} (\widetilde{\nu}_{\rm{L}} -\widetilde{\nu}_{\rm{R}}) 
			\int\limits_{\mu_{\rm{R}}}^{\mu_{\rm{L}}} d\omega \ \frac{1}{N(\omega)^{2}} \Bigg].\nonumber \\
		\end{eqnarray}
	\end{widetext}
	With the Assumption $\widetilde{\nu}_{\rm{R}} = - \widetilde{\nu}_{\rm{L}}$ we get: 
	\begin{align}
	N(\omega)& =  \omega^{4} + 2 A \omega^{2} + B^{2}, \hspace{0.5cm} \nonumber \\ \mbox{with} \
		A & = -(\widetilde{\nu}_{\rm{L}}^{2} + \left|T_{c}\right|^{2} -  \Gamma^{2}/16), \nonumber \\
		B & = \hspace{0.4cm}  \widetilde{\nu}_{\rm{L}}^{2} + \left|T_{c}\right|^{2} +  \Gamma^{2}/16 , 
	\end{align}
	and the integral in equation \eqref{eq.:frictionDD} is given trough $\mathcal I_4(A,B)$ in section \ref{sec.:DQD_aux_int}.
	\subsection{\label{sec.:DQD_fix_point} Fixed point analysis for the double dot}
	The fixed points of a nonlinear two dimensional system can be investigated with standard methods for linear dynamical system \cite{Strogatz2000}.\\
	The general solution for a two dimensional linear system $\dot{\textbf{x}} = A \ \textbf{x}$ 
	is
	\begin{equation}		\label{eq.:linear_solution}
		\textbf{x}(t) = c_{1} e^{\lambda_{1} t} \textbf{v}_{1} + c_{2} e^{\lambda_{2} t} \textbf{v}_{2}
	\end{equation}
	and so determined by the eigenvalues $\lambda_{1,2}$ of the matrix A. 
	The constants $c_{1,2}$ depend on the initial conditions and $\textbf{v}_{1,2}$ are eigenvectors.\\
	The eigenvalues can be obtained from 
	\begin{equation}		\label{eq.:eigenvalues}
		\lambda_{1,2} = \frac{1}{2} (\tau \pm \sqrt{\tau^{2} - 4 \Delta} ),
	\end{equation} 
	thereby $\tau$ corresponds to the trace and $\Delta$ to the determinant of A. 
	These two qualities determine the evolution of the trajectories in the phase plane.\\
	For a fixed point $\textbf{x}^{\ast} $ the condition $\dot{\textbf{x}} = 0$ must be fulfilled. 
	Various classes of fixed points exist, whereas 
	the determinant $\Delta$ decides which kind of point appears. 
	In case of saddle points the determinant is negative and it is positive for spirals or nodes. 
	The difference between a spiral and a node is that for the second one the eigenvalues have no imaginary part.\\
	The trace $\tau$ defines the stability of nodes and spirals, 
	this is caused by the fact that $\tau$ determines the sign of the eigenvalue's real part, for instance 
	with negative real part decaying oscillations occur and the fixed point is stable, see equation \eqref{eq.:linear_solution}.
	There also exist some borderline cases, whereas the centers are the most significant ones, 
	they occur when the trace is equal to zero.\\[2mm] 
	This analysis can be assigned to a two dimensional nonlinear system $\dot{\textbf{x}} = f(\textbf{x})$. 
	By assuming a small disturbance $\textbf{u} = \textbf{x} - \textbf{x}^{\ast}$ from a fixed point, we can
	invest if this disturbance grows or decays by performing a Taylor expansion 
	\begin{eqnarray}
		\dot{u}_{1} &=& f_{1}(x_{1}^{\ast},x_{2}^{\ast}) + u_{1} \frac{\partial f_{1}}{\partial x_{1}} \big|_{x_{1}^{\ast}} 
		+ u_{2} \frac{\partial f_{1}}{\partial x_{2}} \big|_{x_{2}^{\ast}}
		+ {\rm h.t.} \nonumber \\
		\dot{u}_{2} &=& f_{2}(x_{1}^{\ast},x_{2}^{\ast}) + u_{1} \frac{\partial f_{2}}{\partial x_{1}} \big|_{x_{1}^{\ast}} 
		+ u_{2} \frac{\partial f_{2}}{\partial x_{2}} \big|_{x_{2}^{\ast}} 
		+{\rm h.t.} \nonumber \\
	\end{eqnarray}
	The first term is zero and higher terms ($\rm h.t.$) can be neglected because the disturbance is small. So we get a  
	linearised system $\dot{\textbf{u}} = J^{\ast} \textbf{u}$, containing the Jacobi matrix $J^{\ast}$ evaluated at 
	the fixed point coordinates. The above explained analysis can be performed for this system. 
	This is valid as long no borderline cases occur, then the higher terms may be more important.

	\begin{widetext}
		\section{\label{sec.:DQD_aux_int} Auxiliary integrals (DQD)}
				  
		\begin{align}
		\mathcal I_1(\mu,A,B)
		&=\int_{-\infty}^\mu d\omega \frac{1}{\omega^4+2 A\omega^2+B^2} \nonumber\\
		&= \frac{1}{2\sqrt{A^2-B^2}}
			\bigg[ 
				+\frac{1}{\sqrt{A-\sqrt{A^2-B^2}}}
				\arctan\big(\frac{\mu}{\sqrt{A-\sqrt{A^2-B^2}}}\big) \nonumber\\
		&\hspace{16ex}-\frac{1}{\sqrt{A+\sqrt{A^2-B^2}}}
				\arctan\big(\frac{\mu}{\sqrt{A+\sqrt{A^2-B^2}}}\big) \nonumber\\
		&\hspace{16ex}+\frac\pi2\bigg(
					\sqrt{\frac{1}{A-\sqrt{A^2-B^2}}} -\sqrt{\frac{1}{A+\sqrt{A^2-B^2}}}
				\bigg)
			\bigg], \nonumber\\[2ex]
		\mathcal I_2(\mu,A,B)
		&=\int_{-\infty}^\mu d\omega \frac{\omega}{\omega^4+2 A\omega^2+B^2} \nonumber\\
		&= \frac{1}{2\sqrt{A^2-B^2}}
			\bigg[ 
				+\frac12 \ln\big(\mu^2+ A-\sqrt{A^2-B^2}\big) \nonumber\\
		&\hspace{16ex}-\frac12 \ln\big(\mu^2+ A+\sqrt{A^2-B^2}\big)
			\bigg], \nonumber\\[2ex]
		\mathcal I_3(\mu,A,B)
		&=\int_{-\infty}^\mu d\omega \frac{\omega^2}{\omega^4+2 A\omega^2+B^2} \nonumber\\
		&= \frac{1}{2\sqrt{A^2-B^2}}
			\bigg[ 
				-\sqrt{A-\sqrt{A^2-B^2}}
				\arctan\big(\frac{\mu}{\sqrt{A-\sqrt{A^2-B^2}}}\big) \nonumber\\
		&\hspace{16ex}+\sqrt{A+\sqrt{A^2-B^2}}
				\arctan\big(\frac{\mu}{\sqrt{A+\sqrt{A^2-B^2}}}\big) \nonumber\\
		&\hspace{16ex}-\frac\pi2\bigg(
					\frac{1}{\sqrt{\frac{1}{A-\sqrt{A^2-B^2}}}} -\frac{1}{\sqrt{\frac{1}{A+\sqrt{A^2-B^2}}}}
				\bigg)
			\bigg],   \nonumber\\[2ex]    
		\mathcal I_4(A,B)
		&=\int d\omega \frac{1}{\left[\omega^{4} + 2 A \omega^{2} + B^{2} \right]^{2}} \nonumber\\
		&= \frac{1}{8 B^{2}(B^{2}-A^{2})}
			\bigg[ 
							\frac{2 \omega (B^{2}-2A^{2}-A \omega^{2})}{B^{2}+2A\omega^{2}+\omega^{4}} \nonumber \\
			&\hspace{16ex} -\frac{(A^{2}-3B^{2}+A \sqrt{A^{2}-B^{2})} \arctan \frac{\omega}{\sqrt{A-\sqrt{A^{2}-B^{2}}}} }
																{\sqrt{A^{2}-B^{2} } \ \sqrt{A-\sqrt{A^{2}-B^{2}}} } \nonumber \\
			&\hspace{16ex} +\frac{(A^{2}-3B^{2} - A \sqrt{A^{2}-B^{2})} \arctan \frac{\omega}{\sqrt{A+\sqrt{A^{2}-B^{2}}}} }
															{\sqrt{A^{2}-B^{2} } \ \sqrt{A+\sqrt{A^{2}-B^{2}}} } 
					\bigg].   \label{eq.aux_definite_integrals}
		\end{align}
		\newpage
	\end{widetext}	
\end{document}